\documentclass[twocolumn,showpacs,preprintnumbers,amsmath,amssymb,prl]{revtex4}

\usepackage{graphicx}
\usepackage{bm}

\begin{document}

\title{Complex Phase Behavior of Repulsive Step Potential Systems in Three Dimensions }

\author{Yu. D. Fomin}
\affiliation{Institute for High Pressure Physics, Russian Academy
of Sciences, Troitsk 142190, Moscow Region, Russia}

\author{Daan Frenkel}
\affiliation{FOM Institute for Atomic and Molecular Physics,
Amsterdam, The Netherlands}

\author{N. V. Gribova}
\affiliation{Institute for High Pressure Physics, Russian Academy
of Sciences, Troitsk 142190, Moscow Region, Russia}

\author{V. N. Ryzhov}
\affiliation{Institute for High Pressure Physics, Russian Academy
of Sciences, Troitsk 142190, Moscow Region, Russia}

\author{S. M. Stishov}
\affiliation{Institute for High Pressure Physics, Russian Academy
of Sciences, Troitsk 142190, Moscow Region, Russia and Los Alamos
National Laboratory, 87545 Los Alamos, NM, USA}

\date{\today}

\begin{abstract}
The comprehensive computer simulation study of the phase diagram
of the repulsive step potential system in three dimensions is
represented. We show that the system with a simple purely
repulsive isotropic potential demonstrates a number of unusual
features. The maxima and minima on the melting curve are found for
some regions of potential parameters. It is shown that the phase
diagram in $\rho-T$ plane includes two isostructural crystalline
parts separated by the disordered phase which is amorphous at low
enough temperatures. Phase diagram in the (P-T) plane shows that
the transition to the amorphous state occurs approximately along
the extrapolated spinodals. Structural FCC-BCC phase transition is
found at high densities.
\end{abstract}

 \maketitle

After the pioneering work by Hemmer and Stell \cite{stell}, where
the soft core potential with an attractive interaction at large
distances was first proposed for the qualitative explanation of
the isostructural phase transitions in materials such as Ce or Cs,
a lot of attention was paid to the investigation of the properties
of the systems with the potentials that have a region of negative
curvature in their repulsive core
\cite{book1,stell1,stell2,RS2002,RS2003,FRT2006,stanley98,
stanley99,stanley2005,stanley2006,
jagla1,jagla2,jagla3,jagla4,young1,young2,fren97,stishov}. These
studies are particularly relevant because the interatomic
potentials of some pure metallic systems, metallic mixtures,
electrolyte and colloidal systems can be approximated by such type
of potentials. The simplest form of the negative curvature
potential is the repulsive step potential which consists of a hard
core plus a finite repulsive shoulder of a larger radius. It is
well known that systems of particles interacting through the pair
potentials which have repulsive shoulders can possess a rich
variety of phase transitions and thermodynamic anomalies,
including water-like anomalies \cite{stanley98}, liquid-liquid
phase transitions \cite{RS2002,RS2003,FRT2006} and isostructural
transitions in the solid region \cite{young1,young2,fren97}. The
features of greatest interest in the experimental phase diagrams
of $Ce$ and $Cs$ are the isostructural solid-solid phase
transitions and the melting temperature minima and maxima. The
authors of Ref. \cite{young1,young2} succeeded in modelling these
features of the phase diagram in two dimensions.

In this paper we present results, based on computer simulations in
three dimensions, of the phase behavior of the system with the
purely repulsive step potential which reveal a surprisingly
complex phase behavior of the system as a function of the
potential parameters and show new features not found in previous
studies.

The repulsive step potential has the form:
\begin{equation}
\Phi (r)=\left\{
\begin{array}{lll}
\infty , & r\leq d \\
\varepsilon , & d <r\leq \sigma  \\
0, & r>\sigma%
\end{array}%
\right.  \label{1}
\end{equation}
and is characterized by the hard core diameter $d$, the width of
the repulsive step  $\sigma$  and its height $\varepsilon$. In the
low temperature $T<<\varepsilon$ and high temperature
$T>>\varepsilon$ limits the system behaves like a simple hard
sphere systems with hard-sphere diameter $\sigma$ or $d$
correspondingly. Schematic phase diagram may be understood in the
following way: the melting line of the hard sphere system has the
form \cite{hoover} $P=cT/\sigma'^3$, where $c \approx 12$ and
$\sigma'$ is a hard sphere diameter. At low and high temperature
limits one has two straight melting lines which correspond to hard
spheres with diameters $\sigma$ and $d$. At higher temperatures
and densities the line corresponding to $\sigma$ should bend to
the line corresponding to the hard spheres with the diameter $d$.
These lines merge at $T \approx \varepsilon$ and, depending on the
magnitude of the ratio $s=\sigma/d$ one can expect more or less
pronounced anomalies in this region. For high values of $s$ one
can expect well pronounced maximum on the resulting melting curve
which should disappear in the limit $s\rightarrow 1$
\cite{stishov}.

\begin{figure}
\includegraphics[width=7cm]{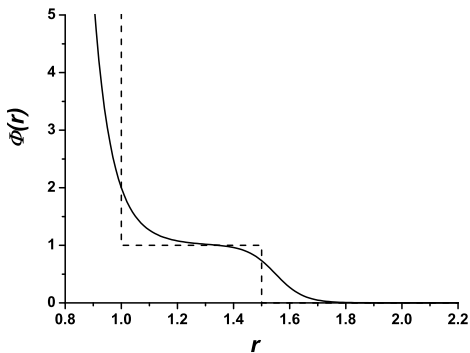}%
\caption{\label{fig:fig1}  A repulsive step potential consisting
of a hard core plus a finite shoulder (dashed line)
($\varepsilon=1, \sigma=1.5$) along with the continuous version of
the potential (\ref{2}) used in the simulations  ($\varepsilon=1,
\sigma_s=1.55$). }
\end{figure}

\begin{figure}
\includegraphics[width=7cm]{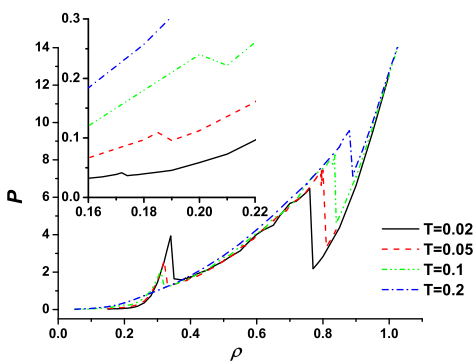}%
\caption{\label{fig:fig2} The family of pressure isotherms for the
purely repulsive step potential (\ref{2}) for $\sigma_s=1.55$. }
\end{figure}

In our simulations we have used a smoothed version of the
repulsive step potential (\ref{1}), which has the form:
\begin{equation}
\Phi (r)=
\left(\frac{d}{r}\right)^{n}+\frac{1}{2}\varepsilon\left(1-\tanh\left(k_0
\left(r-\sigma_s \right)\right)\right) \label{2}
\end{equation}
where $n=14,k_0=10$. We have considered the following values of
$\sigma_s$: $\sigma_s=1.55, 1.35, 1.15$. Further in this paper we
use the dimensionless quantities: $\tilde{{\bf r}}={\bf r}/d$,
$\tilde{P}=P d
^{3}/\varepsilon ,$ $\tilde{V}=V/N d^{3}=1/\tilde{\rho},$ $\tilde{T}%
=k_{B}T/\varepsilon $, omitting the tilde marks.

In Fig.~\ref{fig:fig1} the repulsive step potential is shown along
with its smooth version which was used in our Monte-Carlo and
molecular dynamics simulations.

In order to investigate phase diagrams of the system with the
potential (\ref{2}) we studied isotherms, isochores, radial
distribution functions (RDF), mean-square displacements (MSD),
positional order parameter (POP) in the framework of the standard
molecular dynamics  in the NVT ensemble with periodic boundary
conditions. Temperature is fixed by rescaling the velocities of
the particles whenever necessary \cite{book_fs}. Pressure is
calculated by a direct evaluation in terms of interparticle
forces. To obtain RDF, MSD and POP, we started from the lattice
structure (FCC or BCC), equilibrated for $5\times 10^6$ cycles and
than measured for $3\times 10^6$ cycles with a time step $\delta
t=5\times 10^{-5}$, or from disordered phase obtained by starting
from the crystal one at high temperature and low density. In some
cases the system demonstrates glassy behavior therefore
equilibration run had to be so long.

\begin{figure}
\includegraphics[width=7cm]{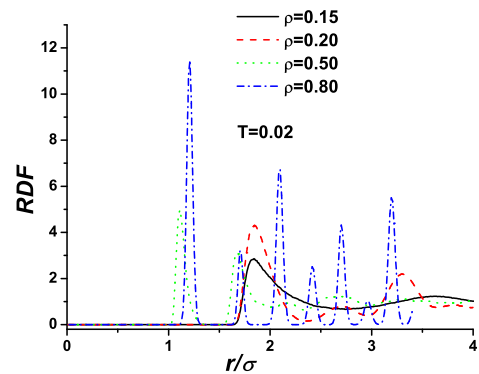}%

\includegraphics[width=7cm]{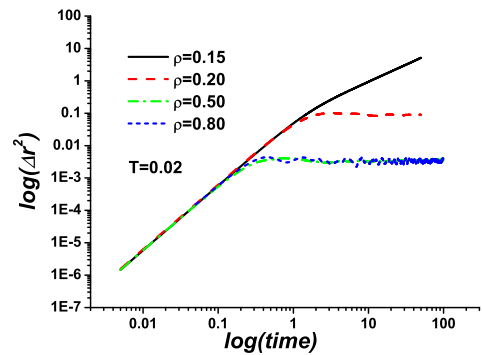}%
\caption{\label{fig:fig3} RDF (upper figure) and MSD for
different densities at $T=0.02$.}
\end{figure}

We also calculated Helmholtz free energy of the system using the
thermodynamic integration in the framework of NVT ensemble Monte
Carlo simulations \cite{book_fs}. For calculating the free
energy of a liquid phase we used integration from dilute gas
limit \cite{book_fs}. Solid phase free energy was calculated by
coupling with Einstein crystal (Frenkel-Ladd method)
\cite{book_fs,6m}. We did $10^6$ equilibration cycles and then
$50000$ production cycles.

In order to locate roughly phase boundaries for crystal phases we
analyzed the behavior of RDF, MSD and POP in the vicinity of van
der Waals loops on the isotherms using MD simulations. In
Fig.~\ref{fig:fig2} the family of pressure isotherms for the
purely repulsive step potential (\ref{2}) with $\sigma_s=1.55$ for
different values of $T$ is shown. One can see that for
$T=0.02,0.05,0.1$ there are 3 sharp bends on the pressure
isotherms and only one sharp bend for $T=0.2$. The nature of the
transitions may be understood from the behavior of RDF and MSD
shown in Fig.~\ref{fig:fig3} for $T=0.02$. It can be seen that for
$\rho=0.15$ (before the first bend on the corresponding isotherm)
RDF and MSD have a liquid-like character. For $\rho=0.2$ RDF and
MSD have solid-like forms. The most interesting behavior
corresponds to the densities in the range between second and third
bends on the isotherms. In this case (see lines corresponding to
$\rho=0.5$) RDF is liquid-like, but MSD has solid-like form. This
means that in the range of densities between second and third
bends on the isotherms at low enough temperatures there exists an
amorphous solid phase, which transforms to the crystalline solid
after the third transition (see RDF and MSD for $\rho=0.8$). In
this density region the crystalline solid is unstable against the
amorphization at all temperatures. The lowest temperature at which
the simulations were done was $T=0.00001$. However, it may be
shown that at high temperatures the system is a liquid (RDF and
MSD are liquid-like). The location of the liquid-glass transition
is a subject of the separate publication. This approach gives the
possibility to obtain the approximate limits of stability
(spinodals) of solid and liquid phases.

\begin{figure}
\includegraphics[width=7cm]{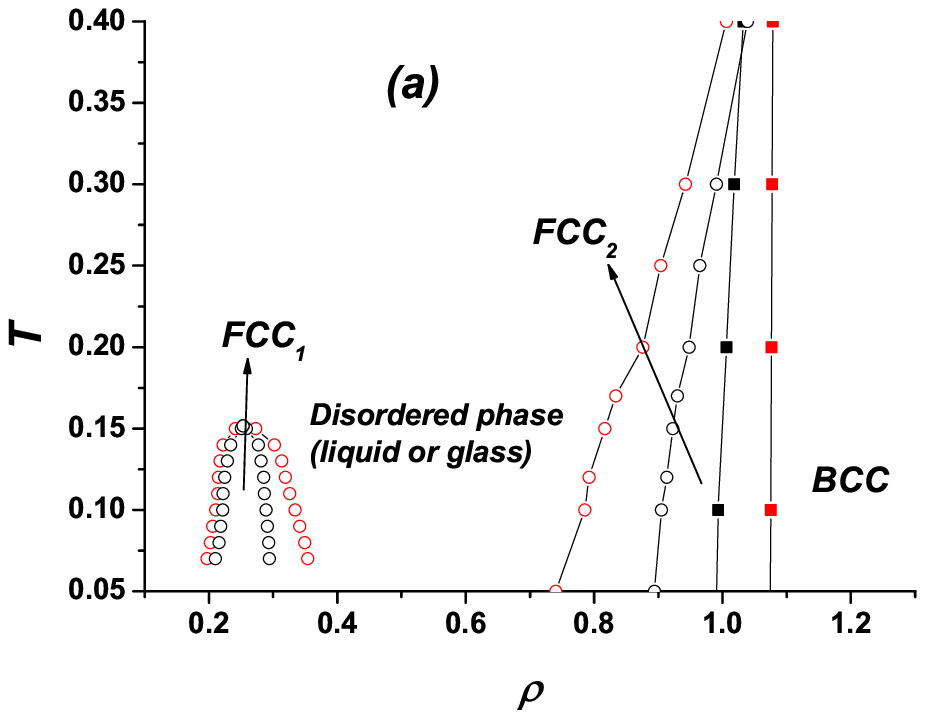}%

\includegraphics[width=7cm]{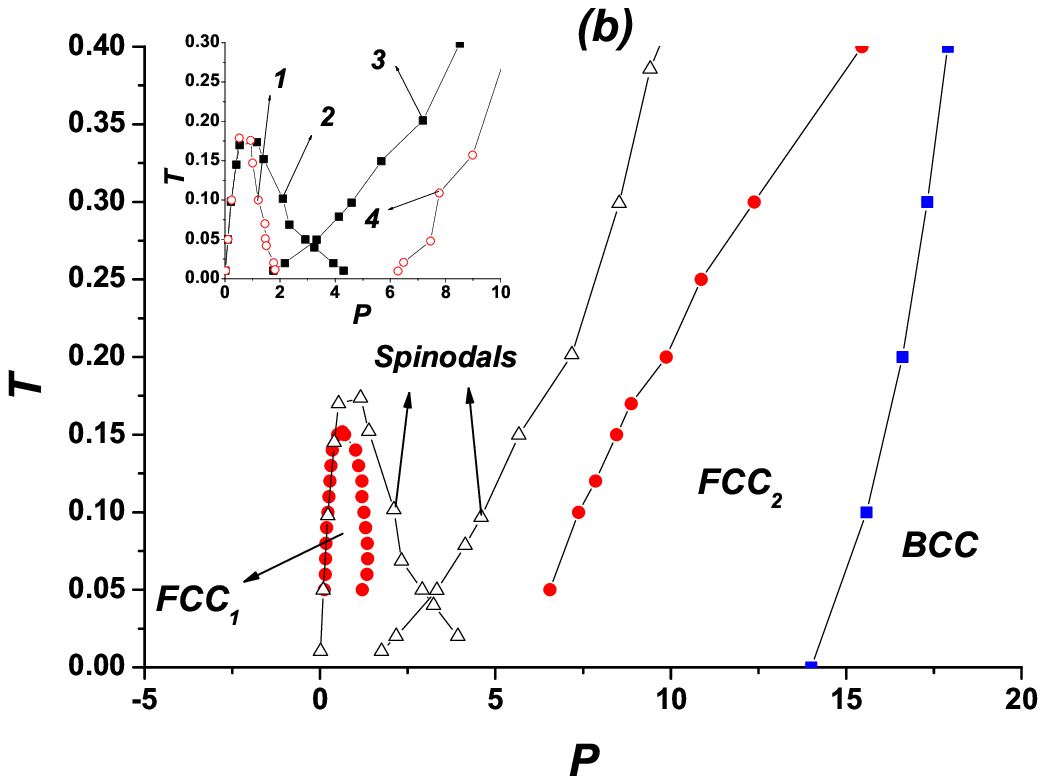}%
\caption{\label{fig:fig4} Phase diagram of the system of particles
interacting through the potential (2) with $\sigma_s=1.55$ in
$\rho-T$ and $P-T$  planes.}
\end{figure}

In Fig.~\ref{fig:fig4}(a) we show the phase diagram of the system
with $\sigma_s=1.55$ as obtained using the double tangent
construction. Open circles determine the boundaries of liquid and
solid (FCC) phases. Squares correspond to the structural FCC-BCC
transition. As it was discussed above there are two FCC phases
separated by the rather wide density range of the disordered phase
which is amorphous at low temperatures. To our knowledge this is a
new type of phase behavior which was not found in previous
studies.

In Fig.~\ref{fig:fig4}(b) the phase diagram of the system with
$\sigma_s=1.55$ is represented in ($T-P$) plane. The circles
correspond to the melting line obtained from the double tangent
construction, and open triangles denote the spinodals obtained
from MD simulations. As one can expect the spinodals lie higher in
temperature than the thermodynamic transition. At higher densities
there is a phase transition between FCC and BCC phases (squares).
The line of the FCC-BCC transition was obtained using the double
tangent construction. MD simulations (RDF and POP behavior)
confirm these results. The thermodynamic melting line (circles)
consists of two parts which should merge at low temperatures.
Unfortunately, we could not calculate the phase diagram between
these parts because of problems with calculation of the Helmholtz
free energy of the disordered state.

In the insert in Fig.~\ref{fig:fig4}(b) the stability line of the
crystalline solid phase is shown by the squares. The open circles
correspond to the stability limits of the disordered phases
obtained from the pressure isotherms (see Fig.~\ref{fig:fig2}).
One can see that the transitions to the amorphous phases occurs
along the extrapolations of the melting lines to the supercooled
region, and the "melting" in this region is nothing but the
pressure induced amorphization
\cite{jagla3,jagla4,mishima,br1,br2}.  It should be noted that due
to the fact that only spinodals are shown in the insert in
Fig.~\ref{fig:fig4}(b), we can not determine whether pressure
induced amorphization is conventional "two-phase melting" or
mechanical instability \cite{jagla3,jagla4,mishima}. In order to
explain the phase diagram shown in the insert in
Fig.~\ref{fig:fig4}(b) let us consider the behavior of the system
under compression at temperatures lower than the melting line
minimum. Taking into account Fig.~\ref{fig:fig2}, it can be seen
that upon increasing pressure one intersects the left hand part of
curve 1 corresponding to the stability line of the liquid phase.
At this line liquid-low-density FCC solid transition occurs. At
the continuation of line 2 the transition from a low-density FCC
solid to amorphous phase takes place. Line 4 corresponds to the
limit of stability of the amorphous phase. At this line amorphous
phase crystallizes and becomes a high-density FCC phase. Upon
decompression from the high-density FCC phase one first meets the
continuation of line 3 where the amorphization occurs. Right hand
part of line 1 corresponds to the limit of stability of the
amorphous phase, where amorphous phase crystallizes to the
low-density FCC phase. Upon further decompression the FCC phase
melts at the left hand part of line 2.

It should be emphasized that we can not reach true thermodynamic
equilibrium state in the density range where we find the amorphous
phase. In principle, the energy of the amorphous phase may be
higher than the energies of FCC$_1$ and FCC$_2$ solids, and in
this case the amorphous phase may exist only because of the very
slow nonergodic kinetics of the system. If the system could be
equilibrated one would find the isostructural FCC$_1$-FCC$_2$
phase transition (see, for example, \cite{fren97}).

\begin{figure}
\includegraphics[width=7cm]{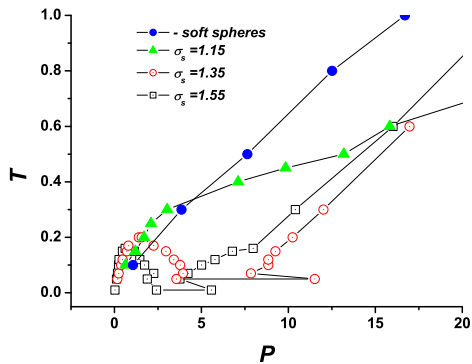}%
\caption{\label{fig:fig5} Phase diagram of the systems of
particles interacting through the potential (\ref{2}) with
$\sigma_s=1.15,1.35,1.55$ in $T-P$ plane. Filled circles
correspond to the melting line of the soft sphere potential
$\Phi(r)=1/r^{14}$ without repulsive shoulder. }
\end{figure}

In Fig.~\ref{fig:fig5} the evolution of the melting line
obtained by equating the Helmholtz free energies is shown as a
function of the size of the repulsive step $\sigma_s$. It can be
seen that the system behaves in qualitative accordance with the
described above model of the mixture of the hard spheres with
two diameters \cite{stishov}. In particular, the melting curve
maximum disappears for $\sigma_s=1.15$. We are going to discuss
this phase diagram in more details in a separate publication.
For the comparison we also show the melting line of the soft
sphere potential $\Phi(r)=1/r^{14}$ without repulsive shoulder
(filled circles).

In conclusion, we performed the comprehensive computer simulation
study of the phase diagram of the system described by the simple
purely repulsive step potential (\ref{2}) in three dimensions and
found a surprisingly complex phase behavior. The maxima and minima
on the melting curve are found for some regions of potential
parameters. It is shown that the evolution of the phase diagram
with change of the parameters may be qualitatively explained in
the framework of the simple model in which the system is
considered as a mixture of two types  of hard spheres. For the
first time it is shown that the phase diagram in $\rho-T$ plane
includes two FCC phases with different densities separated by the
disordered phase which is amorphous at low enough temperatures. We
show that the transition to the amorphous state occurs
approximately along the extrapolated spinodals in the (P-T) plane.
Structural FCC-BCC phase transition is found at high densities.

\begin{acknowledgments}
We thank V. V. Brazhkin for stimulating discussions. The work
was supported in part by the Russian Foundation for Basic
Research (Grants No 05-02-17280 and No 05-02-17621) and NWO-RFBR
Grant No 047.016.001.
\end{acknowledgments}


\end{document}